# A Theory of Enzyme Chemotaxis: Comparison Between Experiment and Model

Farzad Mohajerani,[1†] Xi Zhao,[2†] Ambika Somasundar,[1] Darrell Velegol,[1*] Ayusman Sen[2*]

[1]Department of Chemical Engineering, The Pennsylvania State University, University Park, PA 16802, USA.

[2]Department of Chemistry, The Pennsylvania State University, University Park, PA 16802, USA.

[†]These authors contributed equally.

*Email: velegol@psu.edu, asen@psu.edu

**ABSTRACT**

Enzymes show two distinct transport behaviors in the presence of their substrates in solution. First, their diffusivity enhances with increasing substrate concentration. In addition, enzymes perform directional motion toward regions with high substrate concentration, termed chemotaxis. While a variety of enzymes has been shown to undergo chemotaxis, there remains a lack of quantitative understanding of the phenomenon. Here, we provide a general expression for the active movement of an enzyme in a concentration gradient of its substrate. The proposed model takes into account both the substrate-binding and catalytic turnover step, as well as the enhanced diffusion effect. We have experimentally measured the chemotaxis of a fast and a slow enzyme: urease under catalytic conditions, and hexokinase for both full catalysis and for simple non-catalytic substrate binding. There is good agreement between the proposed model and the experiments. The model is general, has no adjustable parameters, and only requires three experimentally defined constants to quantify chemotaxis: enzyme-substrate binding affinity ($K_d$), Michaelis-Menten constant ($K_M$) and level of diffusion enhancement in the associated substrate ($\alpha$).

## INTRODUCTION

Substrate-induced motility of enzymes has been widely studied in the past decade due to their relevance to the stochastic motion of cytoplasm, the organization of metabolons and signaling complexes, and the convective transport of fluid in cells[1–3]. In addition, anchored enzymes provide a biocompatible power source for the movement of nano and microparticles *in* and *exo vivo* [4–6]. In the presence of their respective substrates, enzymes exhibit enhanced diffusion, likely due conformational fluctuations arising from binding-unbinding of the substrate[7–10]. A second and perhaps more significant observation is their directed stochastic motion towards regions of higher substrate concentration, a phenomenon termed chemotaxis[11–15]. Average enzyme chemotactic velocities can be ~0.2μm/s, more than 10 times of the size of a typical enzymes[12,13]. This behavior has been exploited for enzyme separation[14] and targeted delivery involving crossing of blood-brain behavior[16].

On the basis of a model initially proposed by Schurr et. al., enzymatic chemotaxis can be explained by the favorable binding of enzyme to its substrate, leading to movement up the substrate concentration gradient[17]. This model of binding-driven chemotaxis has been extended and applied to both enzymatic, as well as non-enzymatic systems[13,18]. However, Agudo et al.[19] and Weistuch and Presse [20] recently proposed another model suggesting that substrate binding may lead to negative chemotaxis while "non-specific" phoretic interactions may results in the movement of enzymes toward higher substrate concentrations[19]. Thus, there is no general agreement on the origin of enzyme chemotaxis.

The present work encompasses experimental and theoretical studies of enzyme chemotaxis. We provide a universal expression for the transport of catalytically-active enzymes in a medium with a gradient of substrate concentration. The proposed model takes into account the substrate-binding and the catalytic steps involved in enzyme reactions, as well as the substrate-induced enhanced diffusion[9]. This model is based upon Michaelis-Menten kinetics, linking binding and catalysis to active transport of species, and is also applicable to enzyme-inhibitor interaction and to enzyme cascade reactions Experimentally, we performed chemotactic assay on urease at different substrate concentrations and compared the experimental values with our proposed model. In addition, we experimentally measured chemotaxis of hexokinase in two modes, substrate-binding and full catalysis. We see greater chemotaxis of hexokinase in full-catalysis mode. The model can distinguish between the two conditions of enzyme chemotaxis and while showing close agreement with the experimental values.

## RESULTS AND DISCUSSION

**Enzyme Catalysis and Michaelis-Menten Kinetics.** Every enzymatic reaction, regardless of mechanistic complexity, can be simplified to two general steps: (i) Reversible binding of the substrate (S) to the active site of the enzyme (E) and (ii) the catalytic reaction of enzyme-substrate complex (ES) to form product (P) (eq. 1) [21,22].

$$E + S \underset{k_{-1}}{\overset{k_1}{\rightleftarrows}} ES \overset{k_{cat}}{\rightarrow} E + P \qquad (1)$$

$k_1$ and $k_{-1}$ are the rate constants for the forward and reverse steps in complex formation and $k_{cat}$ is the reaction rate for substrate production. We assume that product formation is an irreversible step, which holds for most of the enzymatic reactions at the early stage when very little product is formed. The complex ES has two possible fates. It can dissociate back to the substrate and free enzyme, or it can transform to product and free enzyme. In either cases, the result is the release of the free enzyme[23] (Figure 1).

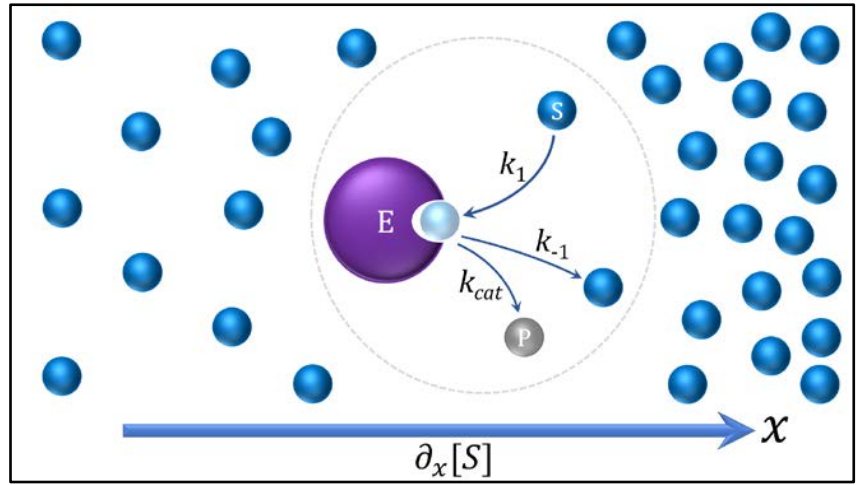

**Figure 1.** Schematic for an enzyme in a concentration gradient of substrate molecules, $\partial_x[S]$. Blue and gray sphere(s) represent substrate and product molecules. After a substrate binds to an enzyme, the resulting complex has two possible fates, unbinding or catalytic turnover. $k_1$, $k_{-1}$ and $k_{cat}$ are the corresponding binding, unbinding and catalytic reaction constants. Binding of the substrate to the enzyme is responsible for the enzymatic chemotaxis.

Under steady-state condition for ES, the Michaelis constant, $K_M$, is defined by eq.2:

$$K_M = \frac{[E][S]}{[ES]} = \frac{k_{-1} + k_{cat}}{k_1} \qquad (2)$$

Since the total enzyme concentration $[E]_T$ is constant ($=[E] + [ES]$), the concentration of enzyme-substrate complex is given by eq. 3:

$$[ES] = [E]_T \frac{[S]}{K_M+[S]} \qquad (3)$$

The reaction velocity (V) is given by eq. 4:

$$V = k_{cat}[E]_T \frac{[S]}{K_M+[S]} \qquad (4)$$

When $[S] \gg K_M$, $[ES] \sim [E]_T$ and the maximum reaction rate $V_{max} = k_{cat}[E]_T$ is achieved. The fraction of free enzyme, $f_E$, can be calculated using eq. 3:

$$f_E = 1 - \frac{[ES]}{[E]_T} = \frac{K_M}{K_M+[S]} \qquad (5)$$

It is also helpful to define $K_d$, the dissociation constant of the ES complex as:

$$K_d = \frac{k_{-1}}{k_1} \qquad (6)$$

$K_d$ is a measure of the enzyme-substrate affinity and is the inverse of binding constant. High $K_d$ indicate weak binding of enzyme to substrate and low $K_d$ indicates strong binding. From eq. 2, if $k_{-1}$ is much greater than $k_{cat}$, $K_M \approx K_d$ [21]. This condition is met for some enzymatic reactions. Also, such a condition is applicable to non-catalytic systems involving binding-unbinding only.

**Theory of Chemotaxis**. Chemotaxis is defined as the directional movement of the enzymes towards high concentrations of substrate molecules[4]. Based on the model proposed by Schurr et al., when probe molecules are placed in a solution with non-uniform concentration of a ligand (i.e. binding molecule), the probe molecules climb up the concentration gradient of the ligand due to favorable binding thermodynamics[17]. This model was modified in Zhao et al. to fit the complex kinetics of hexokinase[13]. On the other hand, the model employed by Zhao et al. does not take into account the enhanced diffusion (i.e., α = 0) and, therefore, systematically underestimates the magnitude of chemotaxis. Here we aim to derive generalized expressions for enzymes following Michaelis-Menten kinetics.

We start from the chemotactic velocity of a single *free* enzyme in a gradient of its substrate. Based on that, we obtain the expression of net velocity of the total enzyme population. According to the derivation provided in Supporting Information, a free enzyme moves toward its substrate with the chemotactic velocity shown below:

$$u_{chem}^E = D_E^0 \frac{\partial_x[S]}{K_d+[S]} \qquad (7)$$

where $u_{chem}$ is the chemotactic velocity of the *free* enzyme, $D_E^0$ is the diffusivity of the enzyme in the absence of substrate, and $[S]$ and $\partial_x[S]$ are the concentration and the concentration gradient of the substrate molecules, respectively. $K_d$, as defined in eq. 6, is the dissociation constant of the enzyme-substrate complex. For the chemotaxis of a single free enzyme towards the substrate, only the *strength of enzyme-substrate binding* is important. The stronger the binding, the smaller the $K_d$ and the larger the chemotactic velocity of the *free enzyme*. The stronger the binding, the smaller the $K_d$ and the larger the chemotactic velocity of the *free enzyme.*

We are interested in the movement of the total enzyme population which consists of both free and bound forms. So, the average chemotactic velocity would be the sum of that for free and bound enzymes with respect to the corresponding fractions:

$$U_{chem}^{net} = f_E \, u_{chem}^{E} + (1 - f_E) \, u_{chem}^{ES} \qquad (8)$$

Where $u_{chem}^{E}$, $u_{chem}^{ES}$ and $U_{chem}^{net}$ are the chemotactic velocity of free, bound, and total enzyme population, respectively. $f_E$ is the fraction of free enzymes given by eq.5. Although free unbound enzyme molecules chemotax due to binding, a substrate-bound enzyme has no thermodynamic reason to chemotax, i.e. $u_{chem}^{ES} \approx 0$. Thus, the chemotactic flux is due to the fraction of free enzyme molecules. The net chemotactic velocity can therefore be formulated as eq. 9:

$$U_{chem}^{net} = f_E u_{chem} = D_E^0 \frac{K_M}{K_M + [S]} \frac{\partial_x [S]}{K_d + [S]} \qquad (9)$$

Eq. 9 shows that both $K_d$ and $K_M$ of an enzyme are determining factors in its chemotactic velocity. Enzyme chemotaxis depends on (i) the strength of enzyme-substrate binding and, (ii) the fraction of free enzyme available for chemotaxis. Low $K_d$ results in high chemotactic velocity of individual free enzyme molecules and high $K_M$ implies that a larger fraction of the enzyme is in the free form and available for active chemotaxis.

Too strong a binding interaction (i.e. irreversible binding) prevents chemotaxis since the fraction of free enzymes quickly goes to zero. Also, as a natural outcome of this model, the presence of inhibitors is expected to decrease enzyme chemotaxis. A competitive inhibitor will block the active site of the enzyme and decrease the fraction of free enzymes that are able to chemotaxis toward the substrate. A non-competitive inhibitor decreases the catalytic turnover rate, $k_{cat}$. A lower turnover rate would also lead to a lower population of free enzymes available for chemotaxis[24]. Both types of inhibition can be accommodated in the proposed model.

If we multiply the chemotactic velocity in eq. 9 by the total enzyme concentration, $[E]_T$, we can obtain the chemotactic flux of the enzymes as following:

$$J_{chem} = D_E^0 \frac{K_M}{K_M + [S]} \frac{[E]_T}{K_d + [S]} \partial_x [S] \qquad (10)$$

In addition to the concentration gradient of substrate, the chemotactic flux of the enzyme depends on $K_d$ and $K_M$, and its diffusion, $D_E^0$. Analogous to diffusive flux, we can define cross-diffusion by equaling the chemotactic flux in eq 10 to $D_{XD}\, \partial_x[S]$, giving [25]:

$$D_{XD} = D_E^0 \frac{K_M}{K_M+[S]} \frac{[E]_T}{K_d+[S]} \qquad (11)$$

In addition to chemotaxis, in the presence of the substrate, the diffusion of enzyme molecules, $D_E$, has been shown to increase from their base/initial diffusion value, $D_E^0$ [1,4,6,9]. At high substrate concentration, when the enzyme is fully bound, its diffusion plateaus to a value that can be attributed to the diffusivity of the ES complex, $D_{ES}$. In moderate substrate concentrations, we can assume the net enzyme diffusivity is an average of the diffusion of the free and bound enzyme molecules with the corresponding diffusions, $D_E$ and $D_{ES}$, respectively. Therefore, we can write:

$$D_E = f_E\, D_E^0 + (1 - f_E)\, D_{ES} \qquad (12)$$

Where $D_E$ is the net enzyme diffusion at any substrate concentration, and $D_E^0$ and $D_{ES}$ are the diffusion of free and bound enzyme molecules, respectively. $f_E$ is the fraction of free enzyme defined in eq. 5. Rewriting this using eq. 5 for the fraction of free enzyme gives the full expression for enzyme diffusion at different substrate concentrations as following:

$$D_E = D_E^0\left(1 + \alpha\, \frac{[S]}{K_M+[S]}\right) \qquad (13)$$

where $D_E^0$ is the diffusion of enzyme in the absence of substrate. $\alpha$ is the enhancement in enzyme diffusivity, $(D_{ES} - D_E^0)/D_E^0$, which is usually 0.3 to 0.8 depending on the type of enzyme. This equation predicts the Michaelis-Menten behavior of enzyme diffusivity that has been observed experimentally in previous reports by our group[9].

To summarize the results on the enzyme diffusion and chemotaxis, we can write down the *total enzyme flux* as the summation of the Fickian diffusion and cross-diffusion which have opposite signs:

$$J = J_{\text{diff}} + J_{\text{chem}} = -D_E\, \partial_x[E]_T + D_{XD}\, \partial_x[S] \qquad (14)$$

Where $D_E$ is the diffusivity of enzyme given by eq. 13, and $D_{XD}$ is the cross-diffusion of enzyme toward substrate given by eq. 11, $\partial_x[E]_T$ is the enzyme concentration gradient and $\partial_x[S]$ is the substrate concentration gradient.

**Experimental Results.** To study the chemotaxis of the enzymes, urease and hexokinase, we used a microfluidic device shown in Figure 2. We tagged urease and hexokinase enzyme by incubating them with excess amount of Alexa Fluor 532 Maleimide and Alexa Fluor 532 NHS Ester, respectively. After purification of the enzyme from the unbound dye molecules, the enzyme concentration is adjusted to 1 $\mu M$ for the microfluidic experiment (supporting Information)[26]. The fluorescently-tagged enzyme solution is pumped through the middle inlet while substrate solution and buffer are pumped in to the bottom and top inlets, respectively. This configuration generates a lateral concentration gradient of substrate across the channel in which the tagged enzymes actively move. Control experiment involved flowing buffer through both side inlets. We avoided premixing the enzyme and substrate to prevent substrate depletion before entering the microfluidic channel. In all the experiments, the flow rate entering each inlet is maintained at 50 µL/hr using a syringe pump. When enzyme intensity profiles are obtained for both experiment and control, the shift of the normalized intensity curve toward the substrate side compared to the control curve is reported as chemotactic shift (inset in Fig 3A). The shift is measured in terms of $\mu m$ at intensity of 0.5. Each set of experiment was done at least 3 times to get a good statistical analysis. (supporting Information).

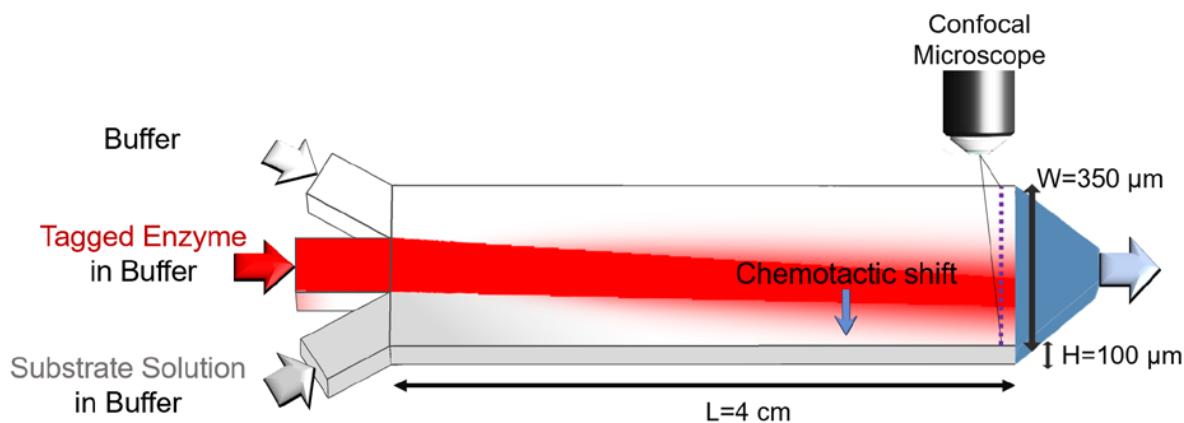

**Figure 2. Schematic of the experimental setup used for measuring enzyme chemotaxis.** A three-inlet one-outlet polydimethylsiloxane (PDMS) microfluidic channel with typical dimensions (L × W × H) of 4 cm × 350 µm × 100 µm was employed. After tagging the enzyme with reactive dye and purifying it, fluorescently-tagged enzyme solution is pumped through the middle inlet while substrate solution and buffer are pumped in to the bottom and top inlets, respectively. This generates a lateral concentration gradient of substrate across the channel in which the tagged enzymes actively move. Control experiment involved flowing buffer through both side inlets. In all the experiments, the flow rate entering each inlet is maintained at 50 µL/hr using a syringe pump. To obtain the intensity profile of the tagged enzyme, confocal scanning microscopy was performed near to the end (~39 mm from the inlet = ~ 34 seconds of interaction time) and around the mid depth of the channel closed to the bottom (~ 25 µm).

**Chemotaxis of Urease.** The first enzyme studied was urease. We did chemotaxis assays of urease at different substrate concentrations, from 10 to 250 mM. Urease is a relatively fast enzyme with $k_{cat} \approx 1.5 \times 10^4 sec^{-1}$ and $K_M \approx 3.3$ mM at 22°C [27]. 100 mM phosphate buffer saline (PBS) was used in all the experiments to maintain the biological pH of 7.2. Tagged-urease at 1 µM concentration enters through the middle inlet. After normalization, the intensity of urease across the channels was plotted (Figure 3A). As can be seen, the urease curve shifts toward the substrate (urea) side. The chemotactic shift of urease is plotted at different concentration of urea injected into one of the side inlets and plotted as a bar graph in Figure 3B. The chemotactic shift increases initially with increasing urea concentration and then plateaus around 150 mM urea. Note that at all concentration of urea, we see a positive chemotaxis shift.

We modeled the chemotaxis of urease with the proposed model summarized in eq. 11 and using the appropriate parameters for urease. A comprehensive simulation was performed by solving the governing equation for urease over the three-dimensional geometry of the channel created in COMSOL Multiphysics software (v5.3) (Figure 1) (Supporting Information)[18,28]. To obtain the highest accuracy, we constructed the comprehensive 3-dimentional geometry of the channel in the software and solved the full fluid field (incompressible Navier-Stokes equation) and mass transport equation (with both convection and diffusion terms) in the domain[29,30]. To do so, two physics of the software has been employed, "Laminar Flow" and "Transport of Diluted Species". The 3-D domain is a long and narrow rectangular channel having 3 inlet and 1 outlet with dimensions similar to the experimentally-used microfluidic device (L × W × H = 4 cm × 350 μm × 100 μm). The details of the modeling is in the Supporting Information. As reported for urease, the diffusion of the enzyme $D_E = 3.1 \times 10^{-11}\ m^2/s$ and the maximum diffusion enhancement at high substrate concentration, $\alpha \approx 30\%$ [7]. For urease, the $k_1$ and $k_{-1}$ are both much higher than the $k_{cat}$ which means that Michaelis-Menten constant is close to dissociation constant of enzyme-substrate complex i.e. $K_M \approx K_d \approx 3.3\ m$M. For the substrate, both the diffusion as well as the consumption rate was considered (Supporting Information). Similar simulations were run for other initial concentrations of urea and the obtained chemotactic shifts are plotted (red curve in Figure 3B). As can be seen, the chemotactic shift is in close quantitative agreement with experimental data.

Another model for enzyme chemotaxis was recently proposed is by Agudo et. al [19]. According to this model, enzyme chemotaxis is the net result of two competing factors, diffusiophoretic velocity due to "non-specific" interactions, as well as velocity arising from specific substrate binding-induced diffusion enhancement. As hypothesized, the phoresis effect will generate positive chemotaxis while enhanced diffusion of enzyme would cause anti-chemotaxis. They proposed anti-chemotaxis below a *critical substrate concentration*, $\rho_S^*$, where phoresis is weak. For urease, they suggested $\rho_S^*$ to be $\sim 30$ mM. Experimentally, we see chemotaxis even at substrate concentration as low as 10 mM $(+1.4 \pm 0.9\ \mu m$ versus $-0.5\ \mu m$ predicted$)$. This model also does not provide a direct tool to estimate chemotaxis of an enzyme since it incorporates an adjustable parameter, the Derjaguin length $(\lambda_e)$, which cannot be determined experimentally[24,31].

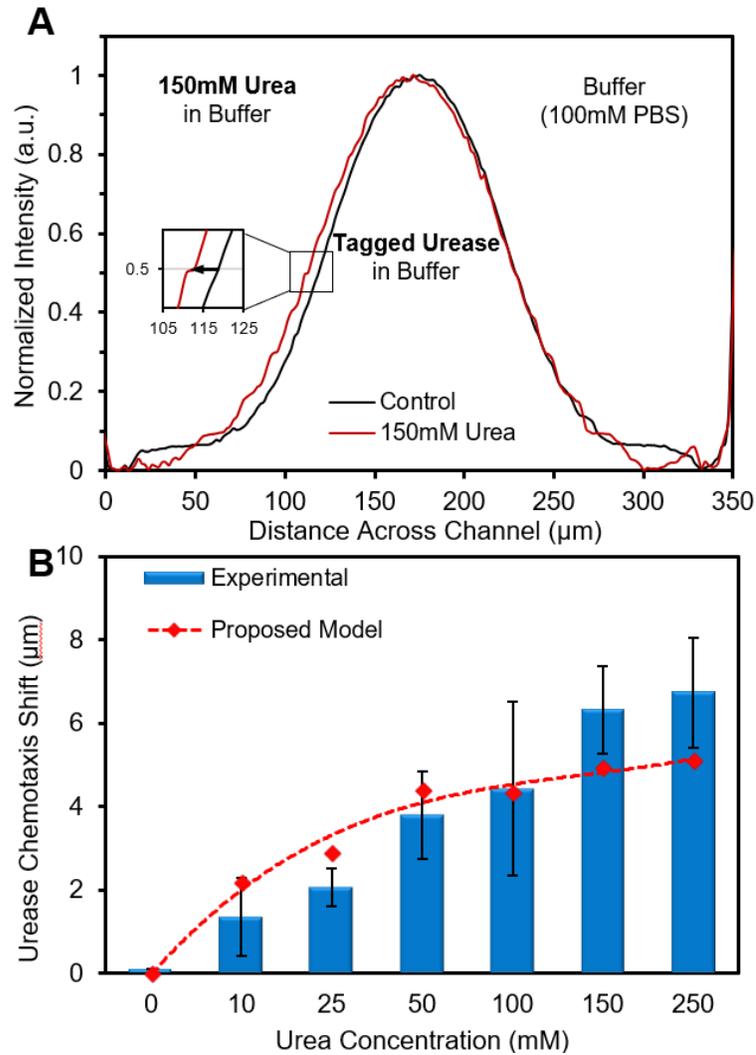

**Figure 3. Chemotaxis of urease enzyme at different concentration gradient of urea. (A)** An example of the normalized intensity profile of tagged urease when the enzyme (1 µM) and its substrate (150 mM) enter through the middle and left inlet, respectively. All the solutions are in 100 mM PBS buffer at pH of 7.2 **(B)** Chemotaxis assay of urease: At different concentrations of urea, experimental values of chemotaxis shift are plotted (bars). The error bars are based on the chemotactic shift calculation for at least 3 experiments and 90% confidence intervals. For each substrate concentration, simulation was run based on the proposed model (equation 14) and the values are plotted as well (dots).

**Chemotaxis of Hexokinase.** To compare the effect of simple substrate binding-unbinding with catalytic turnover in enzyme chemotaxis, we chose hexokinase. In the presence of adenosine triphosphate (ATP) and D-glucose, hexokinase catalyzes the first step in glycolysis metabolic pathway which is the transfer of a phosphoryl group from ATP to D-glucose [32,33]. In the absence of ATP, however, the enzyme simply binds and unbinds D-glucose without catalysis. We examined the chemotaxis of yeast hexokinase from Saccharomyces cerevisiae under both of these conditions, glucose-binding only, as well as full-catalysis. In the former case, fluorescently-tagged hexokinase enters from the middle inlet while solution of 10 mM D-glucose is pumped from one of the side inlets. For the full-catalysis condition, to ensue catalytic turnover step, 10 mM ATP and 20 mM MgCl₂ were added everywhere (in all three inlets). Hexokinase chemotaxed toward D-glucose under both conditions, but higher in full-catalysis mode ($6.2 \pm 2.3\ \mu m$ versus $2.5 \pm 1.1\ \mu m$). In the full catalysis mode, enzyme is freed up from ES complex through both substrate unbinding and product formation with corresponding rates of $k_{-1} \approx 60\ s^{-1}$ and $k_{cat} \approx 200\ s^{-1}$, respectively [34,35]. Thus, compared to only substrate-binding condition, a greater fraction of total enzyme is freed up and available for chemotaxis.

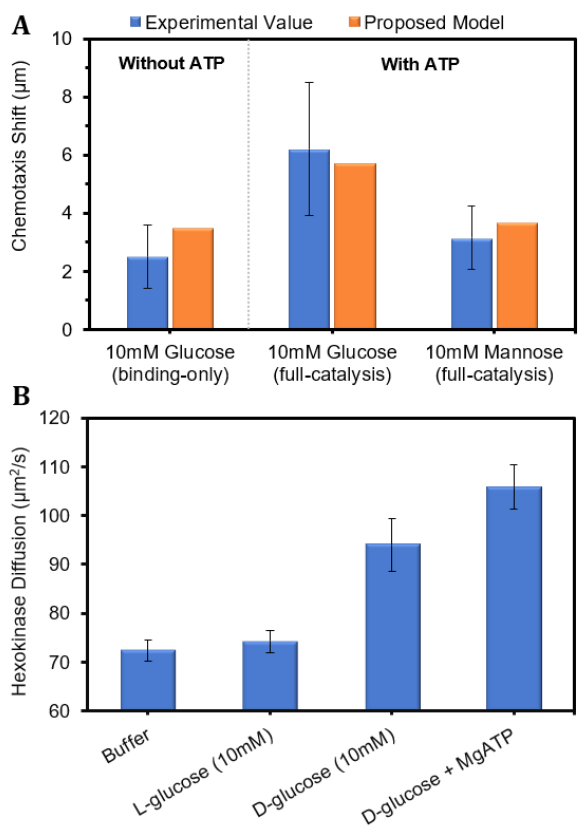

**Figure 4. Chemotaxis and enhanced diffusion of hexokinase under substrate-binding and full-catalysis conditions.** All the experiments were done in HEPES buffer at 50 mM concentration and 7.4 pH. Enzyme is pumped through the middle inlet at 1 μM concentration. **(A)** The chemotaxis shift of the enzyme obtained experimentally under both substrate-binding (left column set) and full-catalysis (right two sets). D-glucose (10 mM) and D-mannose (10 mM) were used as substrates. For the full-catalysis mode (right two sets), ATP (10 mM) and MgCl (10 mM) were added to the solutions in all three inlets. Error bars are 90% confidence intervals obtained based on at least three experiments. The simulated value of chemotaxis based on the proposed model

(eq 14) is also plotted (orange columns in each set). The model shows very close agreement with the experiment. **(B)** The diffusion of fluorescent labeled hexokinase was measured using FCS in the various solutions with or without cofactor and substrate. Hexokinase diffusion increases upon binding/unbinding with D-glucose without catalysis (30.7 ± 5.4 %), but increases further upon full-catalysis (46.2 ± 4.5 %) when both D-glucose and MgATP$^{2+}$ are present. Error bars represent standard deviation of 5 measurements.

To further assess the effect of enzyme-substrate affinity on chemotaxis, we measured the chemotaxis of hexokinase under full-catalysis mode toward one more substrate, D-mannose. Compared to D-glucose, D-mannose binds weaker to hexokinase (higher $K_d$), however, it saturated the enzyme faster (lower $K_M$)[36,37]. Both of these factors are unfavorable for higher enzyme chemotaxis. Experimentally, we observed that hexokinase chemotaxes towards D-mannose ~50% less than D-glucose ($3.2 \pm 1.1 \ \mu m$ versus $6.2 \pm 2.3 \ \mu m$).

A comparison of experimental values of chemotaxis with the modeling values for hexokinase chemotaxis is shown in Figure 4A. For each set, simulation was run using the proposed model with the corresponding parameters (Table S2). For all three cases, the simulation values are very close to the experimental data. For D-glucose, the model predicts that the value of chemotaxis in full-catalysis is higher than the substrate binding mode, which is in agreement with the experiment. On the other hand, when comparing the hexokinase chemotaxis toward D-mannose versus D-glucose in full catalysis mode, the model predicts 40% less chemotaxis toward D-mannose, again close to the 50% less chemotaxis observed experimentally. For comparison, the chemotaxis shift values obtained based on the two previous models by Schurr[17] and Agudo[19] are reported in Table S3. The models significantly over- and underestimate the chemotaxis shift, respectively. The model by Zhao et al. gives results closer to the experimental values but, as reported, still underestimates chemotaxis.[13]

We also measured the diffusion coefficient of hexokinase under different conditions. The measurements of hexokinase diffusion were carried out using 20 nM fluorescent labeled enzymes in various solutions. The diffusion of hexokinase in buffer solution was measured as $7.24 \pm 0.21 \times 10^{-11} m^2/s$, which was close to the theoretical value calculated through Stoke-Einstein equation with a known radius of hexokinase[38]. The significant enhancement in diffusion was observed upon addition of D-glucose (10 mM), (30.7 ± 5.4% enhancement) and even more in presence of D-glucose and MgATP$^{2+}$ (46.2 ± 4.5 % enhancement). Note that, upon addition of L-glucose, no diffusion enhancement was observed confirming that specific binding is crucial for the diffusion enhancement of the enzyme[9].

**Effect of $k_{cat}$ on Chemotaxis.** According to our model, supported by the experiments, the catalytic step enhances chemotaxis of an enzyme in two ways. First, it is another pathway for the release of free enzyme in addition to substrate unbinding. So, the fraction of free enzyme available for chemotaxis, $f_E$, is higher when catalysis occurs. Similar effect was observed for the enzyme diffusion as well. Moreover, the catalytic step of enzymatic reaction results in consumption of substrate in the regions where the enzyme is present. The higher the enzyme population, the faster the substrate consumption rate. This local consumption of substrate leads to a sharper substrate concentration gradient, causing an effect that can be called self-chemotaxis. Therefore, through their self-generated gradient of substrate, enzyme molecules can collectively chemotax faster toward regions of high substrate concentration.

**CONCLUSION**

We suggest that the origin of enzyme chemotaxis is the *binding interaction* between the substrate and the active site on the enzyme while Van der Waals or other "non-specific" interactions play only a minor role. Enzyme chemotaxis is as substrate-specific as enzyme activity (lock and key model)[39]. For example, Zhao et al[13] showed that hexokinase chemotaxes toward D-glucose but not L-glucose, although they are similar in chemical structure and should have the same long-ranged non-specific interactions[13]. Moreover, other studies also showed the directional movement of molecules in binding-driven systems, confirming the importance of binding for molecular chemotaxis[18,40–42].

The co-localization of the enzyme and the substrate (i.e., chemotaxis) lowers the chemical potential of the system due to favorable binding. Part of the thermodynamic driving force arises from the entropically favored expulsion of water molecules from the enzyme pocket upon binding. Additional flows can result from binding and catalysis-induced conformational changes of the enzyme. As suggested, these hydrodynamic flows can propel active and passive particles in solution[19,43–46].

In summary, we have developed a general model for the chemotaxis of enzymes in their substrate concentration gradient which relies only on experimentally measured kinetic parameters for the enzyme, and has no adjustable parameters. It takes into account both steps in any enzymatic reactions, substrate binding and catalytic turnover. The proposed model was tested on chemotactic movement of two different enzymes, urease and hexokinase. We showed that catalytic turnover step has an enhancing effect on both enzyme's diffusion and chemotaxis, the effect which is well-captured by the model quantitatively. The model, which is based on Michaelis-Menten kinetics, is general and can be also applied to other systems: e.g., systems involving inhibitors and enzyme cascade reactions where the model links the reaction system to active transport of species.

**AUTHOR INFORMATION**

**Corresponding Authors**

Darrell Velegol (velegol@psu.edu) and A. Sen (asen@psu.edu).

**Notes.** The authors declare no competing interests.

ACKNOWLEDGEMENTS

We thank Dr. Henry Hess, Steve Granick, Ramin Golestanian, and Jaime Agudo-Canalejo for insightful discussions. The work was supported by Penn State MRSEC funded by the National Science Foundation (DMR-1420620) and by NSF CBET-1603716.

Supporting Information

**A Theory of Enzyme Chemotaxis: Comparison Between Experiment and Model**

Farzad Mohajerani,[1†] Xi Zhao,[2†] Ambika Somasundar,[1] Darrell Velegol,[1*] Ayusman Sen[2*]

[1]Department of Chemical Engineering, The Pennsylvania State University, University Park, PA 16802, USA.

[2]Department of Chemistry, The Pennsylvania State University, University Park, PA 16802, USA.

[†]These authors contributed equally.

*Email: velegol@engr.psu.edu, asen@psu.edu

## 1. Tagging Enzymes

Urease (from Jack Bean, TCI chemicals) was tagged with thiol reactive Alexa Fluor 532 labeling kit (Thermo Fisher Scientific). 1 mg urease was mixed with fluorescent dye in the vial with 0.5 ml 100 mM PBS buffer at 4 °C for 3 hours. The vial was covered with aluminum foil to prevent exposure to light. The fluorescent labeled enzymes were purified by applying P-30 Fine size exclusion purification column with 100 mM PBS buffer[1]. The concentration and tagging ratio of dye-enzyme complexes were determined by UV-Vis. The dye ratio of tagged urease is 1.1. Similar procedure as tagging urease, Hexokinase (from Saccharomyces cerevisiae; Sigma-Aldrich) were fluorescently labeled with amine reactive Alexa Fluor 488 (protein labeling kit, Thermo Fisher Scientific) for microfluidic experiments and Alexa Fluor 532(protein labeling kit, Thermo Fisher Scientific) for diffusion measurements under FCS. 2.5 mg HK was dissolved in 50 mM pH = 7.0 HEPES buffer with one vial of fluorescent dye the kit provides and 10 mM mannose. The labeling reaction was carried out in ice bath wrapped with aluminum foil for 3 hours. Following instruction, the column was assembled with P-30 Fine size exclusion purification resin and was flowed 50 mM HEPES buffer pH = 7.4. The enzyme-dye complexes were separated from free dye molecules by passing solution through purification column. The dye per enzyme molecules was finally around 1.

## 2. Microfluidic Experiments

A three-inlet one-outlet polydimethylsiloxane (PDMS) microfluidic channel with dimensions 4 cm (L) x 350 (W) μm x 100 μm (H) was used to measure the chemotaxis of fluorescently-tagged enzymes under confocal laser scanning microscopy. The shifting mode of chemotaxis was used, where the fluorescently tagged enzymes were flown through the middle channel, and substrate and buffer were flown in each of the side channels respectively, at a particular rate of 50 $\mu l/hr$. We flow the substrates and the enzymes through separate inlets of the microfluidic channel to prevent pre-mixing before the experiment, as pre-mixing could lead to partial or complete depletion of substrate. By flowing the substrate and enzymes separately, we can carefully maintain the interaction time between substrate and enzyme by their residence time in the channel. The flow profile of the fluorescently-tagged enzymes with and without substrate was obtained by recording the fluorescent intensity profile across the width (350 µm) of the microfluidic channel. Optical scans were taken across the channel width at a distance of 3.8 cm from the inlet (~34 seconds residence time) using the confocal laser scanning microscope. The chemotactic shift of the enzymes was measured perpendicular to the direction of flow.

### 3. Microfluidic Experiment Analyses

The recorded videos were each taken for 5 minutes after the flow becomes fully stable in the channel. The average over each five-minute video is considered as 1 experiment. Typically, the number of frames in each video is approximately 680 frames. The stability of the flow was ensured by making sure that there are no debris or bubbles in the microfluidic inlets and entire length of the channel before the start of recording. The flow rate of 50 µl/ hr maintains the residence time of the enzyme and substrate at 34 seconds. The analysis of the fluorescent intensity profile was done using Image J. We take the average intensity of each five-minute video and compared the experiment with the control case i.e. in the absence of substrate. The chemotactic shift was calculated by subtracting the distance of the control from the distance of the experiment at a normalized fluorescence intensity of 0.5 on the side where the substrate is introduced (Figure 2A in the main manuscript).

### 4. Derivation of an Expression for Chemotaxis of Free Enzymes.

Here we want to find the expression linking the enzyme movement to the binding of a substrate in a substrate concentration gradient, using the equation: $E + S \rightleftarrows ES$. This preferential movement, chemotaxis, is due to the fact that the enzyme-substrate binding is a thermodynamically favorable event.

In order to observe and follow the changes, we consider a control volume, $V$. At equilibrium, this finite volume incorporates $n_E, n, n_S$ and $n_w$ as the number of molecules of free enzyme, enzyme-substrate complex, substrate and water/solvent (Figure S1). We assume that the population of substrate is much higher than that of enzyme and ES complex which is a valid assumption. Next step, we want to calculate the free energy change of the system, $\delta G$, upon adding of $\mathcal{E}_E$ molecule of free enzyme to the box. Upon addition of $dn_E$, in order to return to equilibrium, the binding reaction moves forward and convert some $E$ to $ES$. We can denote the change as $\delta \xi$. The entire hypothetical experiment is shown in Figure S1.

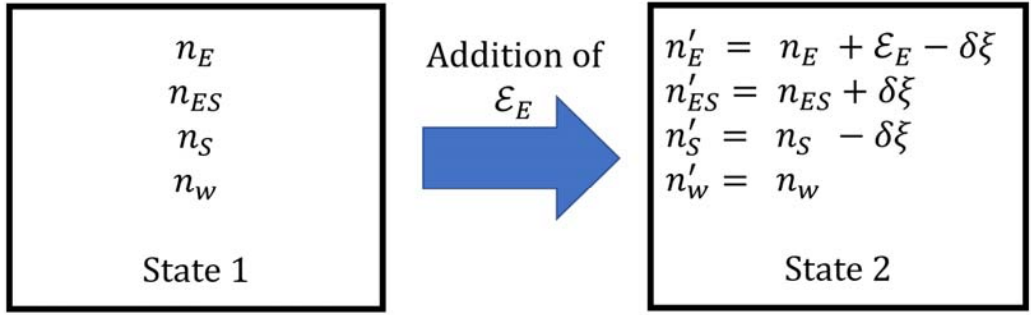

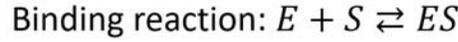

**Figure S1.** Schematic of the finite control volume we assumed in order to obtain the change in the total Gibb's free energy of the system, $\delta G$, upon addition of small number of enzyme molecules, $\mathcal{E}_E$. Initially, the system is at equilibrium with composition of $n_i$. Upon addition of $\mathcal{E}_E$ number of free enzyme molecules, the system reaches equilibrium with a new composition, $n_i'$.

We can write the change in the total free energy of the system (Figure S1). The change from system 1 to system 2 in total free energy (G) at constant temprature and pressure is for small changes[2]:

$$\delta G = \sum_i \mathcal{E}_E \mu_i \qquad\qquad i = \{E, ES, S\} \quad \text{(S1)}$$

Where $\mu_i$ and $\mathcal{E}_i$ are the chemical potential and change in number of molecules of each species. The chemical potential for species $i$ can be defined as:

Solute: $\qquad\qquad \mu_i = \mu_i^0 + kT \ln(\gamma_i c_i / c^0) \qquad\qquad i = \{E, ES, S\} \quad \text{(S2)}$

$\gamma_i$ is the activity coefficient and $\mu_i^0$ is the standard state chemical potential at standard concentration, $c^0$, which is usually 1 M. $c_i \equiv n_i/V$ is the concentration of species $i$ with $n_i$ number of molecules in the $V$ control volume. While for water $\gamma_W \approx 1$ since it is nearly pure, for the other species in general the $\gamma_i \neq 1$ since they are dilute. However, since E, S, and ES are quite dilute, their infinite dilution activity coefficients are applied, and these will be very close to constant for the entire system. , the chemical potential of species i will be a function of only the concentration of species i.

After the addition of small number of enzyme molecules, $\mathcal{E}_E$, the extent of binding reaction changes by $\delta\xi$ number of molecules. Therefore, the change in total free energy of the system from state 1 to state 2 is[2]:

$$\delta G = (\mathcal{E}_E - \delta\xi)\,\mu_E(c_E) + \delta\xi\,\mu_{ES}(c_{ES}) - \delta\xi\,\mu_E(c_S) \qquad\qquad \text{(S3)}$$

After regrouping we obtain:

$$\delta G = \mathcal{E}_E\,\mu_E(n_i) + \delta\xi\,(\mu_{ES}(c_{ES}) - \mu_E(c_E) - \mu_S(c_S)) \qquad\qquad \text{(S4)}$$

We know at equilibrium, Gibbs-free energy of the system is at minimum with respect to reaction coordinate, $\partial G/\partial \xi = \sum_i v_i \mu_i = \Delta G_{rxn}^0 + kT \ln \frac{c_{ES}/c_{ES}^0}{c_E/c_E^0 \cdot c_S/c_S^0} \frac{\gamma_{ES}}{\gamma_E \cdot \gamma_S} = 0$, where $\Delta G_{rxn}^0$ is the standard Gibbs-energy change of reaction and $\gamma_i$ are the infinite dilution activity coefficients [3]. So, the last term in the equation above is zero for small $\delta\xi$. Assuming that the $\gamma_i$ are approximately constant and using that the $c^0 = 1$ M, we define the concentration-based (apparent) dissociation constant for this reaction as $K_d = \frac{c_E \cdot c_S}{c_{ES}}$, with unit of molar ($M$). Therefore, at (or near) equilibrium, equation (S4) can be simplified as below:

$$\delta G = \mu_E(c_E) \, \mathcal{E}_E \tag{S5}$$

Now the chemical potential of free enzyme is

$$\mu_E \equiv \lim_{\mathcal{E}_E \to 0} \delta G / \mathcal{E}_E = \mu_E^0 + kT \ln(\gamma_E c_E / c_E^0) \tag{S6}$$

For the next step, we need to calculate the thermodynamic force on a free enzyme molecule arising from substrate binding. To do so, we need to calculate the change in the *enzyme chemical potential, $\delta\mu_E$,* when $\mathcal{E}_E$ number of enzyme molecules move from one region with substrate concentration of $c_S$ to the neighboring region with slightly higher concentration of substrate, $c_S + \delta c_S$ (Figure S2).

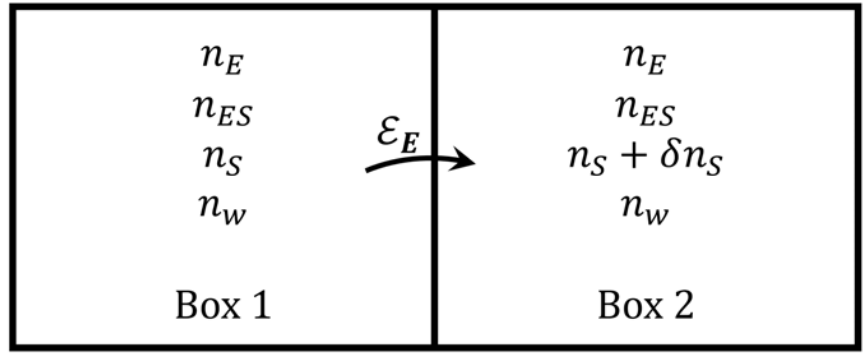

**Figure S2.** Schematic of two neighboring finite control volume with identical volume of $V$ at local equilibrium. A small number of enzyme molecules, $\mathcal{E}_E$, moves from left box to right box where the population (concentration) of substrate is $\delta n_S$ ($\delta n_S/V$) higher.

To do so, first we reconfigure equation S6. Based on the experimental condition we can assume that the concentration of substrate is much more than the that of enzyme and ES complex, $c_S \gg c_E, c_{ES}$. From the initial equilibrium condition, the fraction of free enzyme ($c_E$) to substrate-bound enzyme ($c_{ES}$) is $c_E/c_{ES} = K_d/c_S$. Also using the definition that $c_E^T = c_E + c_{ES}$, we obtain the concentration of free enzymes as a function of substrate concentration:

$$c_E = c_E^T \frac{K_d}{c_S + K_d} \tag{S7}$$

Where $K_d$ is the apparent dissociation constant of the binding reaction. Now, if we plug in the equation S7 in to equation S6, we obtain:

$$\mu_E = \mu_E^0 + kT \ln\left(\gamma_E \frac{c_E^T}{c_E^0} \frac{K_d}{c_S + K_d}\right) \tag{S8}$$

Therefore, we obtained an expression that links *chemical potential of free enzymes* to *substrate concentration*, $c_S$. Using this, we can easily find the chemotactic movement of enzyme driven by substrate. To do so, we need to calculate the change in the free enzyme chemical potential between the two boxes depicted in figure S2 that have different $c_S$. This difference can be written as:

$$\mu_E^2 - \mu_E^1 = \delta\mu_E = \frac{\partial \mu_E}{\partial c_S} \delta c_S + \sigma(\delta c_S^2) \tag{S9}$$

Neglecting higher order, $\sigma(\delta c_S^2)$, and taking the partial derivative of $\mu_E$ (from Eq. S8) with respect to $c_S$ give:

$$\delta\mu_E \approx -kT \frac{1}{c_S+K_d} \delta c_S \tag{S10}$$

Therefore, using the above equation, the thermodynamic force on the free enzyme molecules arising from substrate binding is:

$$F_{binding} = -\lim_{\delta x \to 0} \frac{\delta \mu_E}{\delta x} = \lim_{\delta x \to 0} \left[ kT \frac{1}{c_S+K_d} \frac{\delta c_S}{\delta x} \right] = kT \frac{1}{c_S+K_d} \frac{\partial c_S}{\partial x} \tag{S11}$$

Knowing the force, the chemotactic velocity of free enzymes can be calculated using the mobility of enzyme molecules, $\mathcal{M} = 1/6\pi\eta R_E = D_E/kT$

$$u_E = \mathcal{M} F_b = D_E \frac{1}{c_S+K_d} \frac{\partial c_S}{\partial x} \tag{S12}$$

where $u_E$ is the chemotactic velocity of enzyme.

### 5. Modeling of Enzyme Chemotaxis

We have performed the simulation on the microfluidic experiments using COMSOL Multiphysics (v5.3). To obtain the highest accuracy, we constructed the comprehensive 3-dimentional geometry of the channel in the software and solved the full fluid field (incompressible Navier-Stokes equation) and mass transport equation (with both convection and diffusion terms) in the domain. To do so, two physics of the software has been employed, "Laminar Flow" and "Transport of Diluted Species". The 3-D domain is a long and narrow rectangular channel having 3 inlet and 1 outlet with dimensions similar to the experimentally-used microfluidic device (L × W × H = 4 cm × 350 μm × 100 μm). (Figure 1 in the main manuscript).

The simulation first obtains the flow pattern in the channel by solving the incompressible Navier–Stokes equation. From each inlet, 50 μl/hr solution is pumped in. Soon after entering the main channel, the flow velocity profile becomes fully-developed and remains constant throughout the channel. The fully-developed velocity profile at a cross-section of the channel, $\mathbf{u}(x,z)$, is shown in Figure S3. The velocity is zero at the walls (no-slip condition) and maximum in the middle. The average velocity is ~1.13 mm/s which give the residence time of ~34 sec for travelling the length of the main channel.

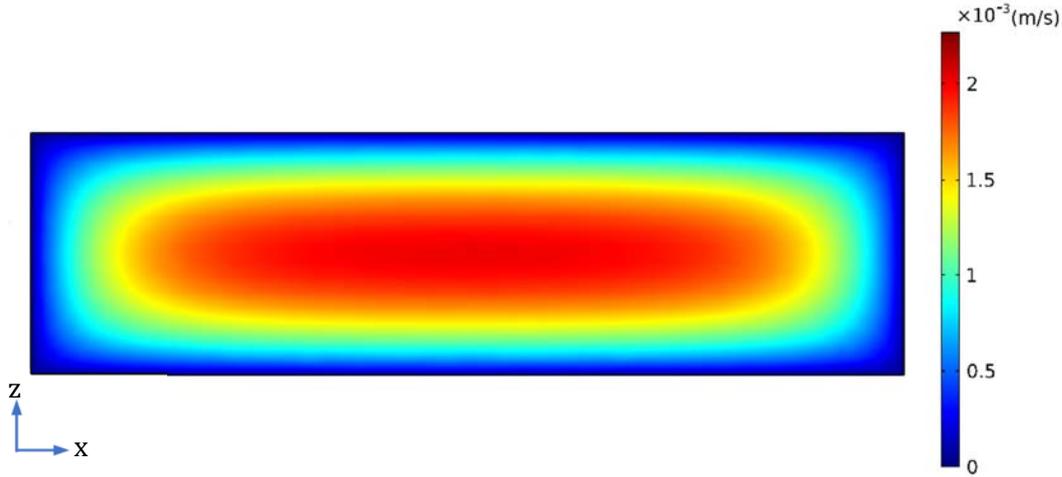

**Figure S3.** The fully developed flow velocity profile, $u(x,z)$ at a cross-sectional plane in the main channel of the microfluidic device used for chemotaxis experiments. The fluid velocity along the channel ($y$ direction) is maximum in the middle and zero along the walls (no-slip condition). Average and maximum velocity along the channel are 1.13 and 2.27 mm/s, respectively.

After obtaining the velocity profile, the simulation solves the two mass transfer equations defined the two species in the system, enzyme and substrate. The governing equation for mass transfer of enzyme and substrate are as following:

For the enzyme:    $\nabla \cdot (-D_E \nabla c_E + D_{XD} \nabla c_S) + \mathbf{u} \cdot \nabla c_E = 0$    (S13)

For the substrate:    $\nabla \cdot (-D_S \nabla c_S) + \mathbf{u} \cdot \nabla c_S = R_S$    (S14)

Which $C_E$ and $C_S$ are the enzyme and substrate concentrations, $D_E$ and $D_S$ are their diffusion coefficients respectively. $D_{XD}$ is the cross-diffusion of the enzyme toward the substrate. $\mathbf{u}$ is the velocity profile in the channel that is solved in the first step of the simulation. $\nabla$ is the gradient operator. Unlike the enzyme, substrate is consumed along the channel at a specific rate, $R_S$, which can be calculated by Michaelis-Menten expression as following:

$$R_S = k_{cat} c_E \frac{c_S}{K_M + c_S}$$    (S15)

Which $k_{cat}$ is the rate of catalytic turnover and $K_M$ is the Michaelis-Menten constant of the enzyme.

Equation S13 is the governing equation for enzyme that has both diffusive ($-D_E \nabla c_E$) and chemotactic terms ($D_{XD} \nabla c_S$). Depending on the model, the appropriate expression for $D_E$ and $D_{XD}$ was used (Table S1). The diffusivity and kinetic parameters values are listed in Table S2 for the enzymes used in this study. For hexokinase, the predictions based on our proposed model along with the two existing models by Schurr et al.[4] and Agudo et al[5] are given in Table S3.

The enzyme enters through the middle inlet and is always kept at 1 µM experimentally as well as in the simulations. The substrate concentration varies for each studied enzyme. For urease, chemotaxis assay was done with urea concentration of 10-250 mM (Figure 3B in the main manuscript). Also, study on hexokinase was done with 1 and 10 mM D-glucose and D-mannose and the values are reported in Table S3 as well as Figure 3A in the main manuscript.

**Table S1**: Diffusion and cross-diffusion (chemotaxis) of the enzyme, based on the different models, used to solve the governing equation S13 for the enzyme. In the equations, $c_E$ and $c_S$ are the enzyme and substrate concentration. $D_E^0$ and $\alpha$ are the base diffusion (in buffer) and the level of enhanced diffusion of the enzyme. Also, $K_d$ and $K_M$ are the enzyme-substrate dissociation constant and Michaelis-Menten constant of the enzyme, respectively. In the model by Agudo et al., $\lambda_e$ is the Derjaguin length; and $R_E$ is the radius of the enzyme which is calculated based on Stokes-Einstein relation form enzyme diffusion, $D_E^0$.

|  | Diffusion of the enzyme ($D_E$) | Cross-diffusion of the enzyme ($D_{XD}$) |
|---|---|---|
| Proposed Model | $D_E^0 \left(1 + \alpha \dfrac{c_S}{K_M + c_S}\right)$ | $D_E^0 \dfrac{K_M}{K_M + c_S} \dfrac{c_E}{K_d + c_S}$ |
| Schurr et al.[4] | $D_E^0$ | $D_E^0 \dfrac{c_E}{K_M + c_S}$ |
| Agudo et al.[5] | $D_E^0 \left(1 + \alpha \dfrac{c_S}{K_M + c_S}\right)$ | $\left[6\pi D_E^0 R_E \lambda_e^2 - \alpha D_E^0 \dfrac{K_M}{(K_M + c_S)^2}\right] c_E$ |

**Table S2.** The corresponding diffusion and kinetic parameters for the modelling of urease and hexokinase at 22°C.

| Enzyme (Substrate) | Diffusion in buffer (m²/s) | Enhanced diffusion ratio ($\alpha$) | $K_M$ (mM) | $K_d$ (mM) | $k_{cat}$ (sec⁻¹) | Substrate Diffusion (m²/s) |
|---|---|---|---|---|---|---|
| **Urease** [6,7] (urea) | $3.1 \times 10^{-11}$ | 0.3 | 3.3 | 3.3 | 15000 | $12 \times 10^{-10}$ |
| **Hexokinase** [8–12] (D-glucose and MgATP²⁺) Full-catalysis | $7.2 \times 10^{-11}$ | 0.35 | 0.120 | 0.025 | 200 | $6.8 \times 10^{-10}$ |
| **Hexokinase**[8–12] (D-mannose and MgATP²⁺) Full-catalysis | $7.2 \times 10^{-11}$ | 0.35* | 0.050 | 0.060 | 145 | $6.8 \times 10^{-10}$ |
| **Hexokinase**[8–12] (D-glucose) Substrate-binding only | $7.2 \times 10^{-11}$ | 0.45 | 0.025** ($= K_d$) | 0.025 | 0 | $6.8 \times 10^{-10}$ |

*Expected value. **Since there is no catalysis, the $K_M$ is set to $K_d$.

**Table S3:** Experimental and modelling values for chemotaxis of hexokinase under different conditions in $\mu m$.

|  | D-glucose (1mM) (Substrate-binding) | D-glucose (10mM) (Substrate-binding) | D-glucose (10mM) + MgATP$^{2+}$ (10mM) (Full-catalysis) | D-mannose (10mM) + MgATP$^{2+}$ (10mM) (Full-catalysis) |
|---|---|---|---|---|
| Experimental Value of Chemotaxis | $3.4 \pm 1.35$ | $3.15 \pm 1.09$ | $6.2 \pm 2.28$ | $2.51 \pm 1.09$ |
| Proposed Model | 4.6 | 3.5 | 5.9 | 3.7 |
| Model by Schurr et al. | 23.2 | 18.0 | 23.5 | 20.5 |
| Model by Agudo et al.* | $-0.15$ | 0.2 | 0.5 | 0.1 |

*Based on the suggested upper limit for Derjaguin length of 3 Å.

## 6. FCS experiments

The fluorescent correlation spectroscopy is a custom-built microscope equipped with time-correlated single-photon counting (TCSPC) as described before[13,14]. The laser light is a PicoTRAIN laser at 532 nm light with 80 MHz frequency. When the laser is on, the light goes through an IX-71 microscope to excite samples. Then, the fluorescent light emitted by samples was filtered by a dichroic beam splitter (Z520RDC-SP-POL, Chroma Technology), focused in a confocal pinhole, a 50 μm, 0.22-NA optical fiber (Thorlabs), pre-amplified by photomultiplier tube(HFAC-26) and then recorded by a time-correlated single-photon counting (TCSPC) board (SPC-630, Becker and Hickl). The signal is fluctuating, for fluorescent molecules diffusing in and out of the confocal plane. The fluctuations are then auto correlated using the equation S16 and S17 and fit by a two-component 3D model to determine the diffusion coefficient of fluorescent molecules[15].

$$G(\tau) = \sum_{i=1}^{n} \frac{1}{N_i} \left[ 1 + \left( \frac{\tau}{\tau_D^i} \right) \right]^{-1} \left[ 1 + \left( \frac{1}{w} \right)^2 \left( \frac{\tau}{\tau_D^i} \right) \right]^{-\frac{1}{2}} \quad \text{(S16)}$$

where
$$\tau_D^i = \frac{r^2}{4D_i} \quad \text{(S17)}$$

Here, $N_i$ is the average number of fluorophores of the $i^{th}$ species in the observation volume, $\tau$ is the auto-correlation time, $w$ is the structure factor, which is defined as the ratio of height to width of the illumination profile, and $\tau_D^i$ is the characteristic diffusion time of the $i^{th}$ fluorescent particle with diffusion coefficient $D_i$ crossing a circular area with radius $r$.

To measuring the diffusion coefficient of hexokinase, the laser power was adjusted to 25 μW and the confocal volume was calibrated before each experiment using 50 nm fluorescent beads in deionized water. Each solution contains 20 nM fluorescent hexokinase. To determine τ$_D$, the auto correlated curves were fit to equation 1, using the Levenberg–Marquardt nonlinear least- squares regression algorithm with Origin software.